\documentclass[conference]{IEEEtran}
\usepackage{color}
\usepackage{amsmath}
\usepackage{graphicx}
\usepackage{listings}
\usepackage{multirow}
\usepackage{scalefnt}
\usepackage{rotating}
\usepackage{enumerate}
\usepackage{hyperref}
\usepackage[tight,footnotesize]{subfigure}
\usepackage{url}
\usepackage{listings}
\usepackage{framed}
\usepackage[symbol]{footmisc}

\newtheorem{mydef}{Definition}
\newtheorem{theorem}{Theorem}
\newtheorem{remark}{Remark}
\newtheorem{corollary}{Corollary}

\newcommand{\code}[1]{\texttt{\footnotesize #1}}
\newcommand{\remove}[1]{}

\begin{document}
\title{Event-Flow Graphs for Efficient Path-Sensitive Analyses}
%\title{Event-Flow Graphs for Efficient Path-Sensitive Analyses\titlenote{Theorems' \& corollaries' proofs are omitted due to space limitation. We intend to include them in the final version. Proofs can be found at~\cite{RESULTS}.}}

\author{\IEEEauthorblockN{Ahmed Tamrawi and Suresh Kothari}
\IEEEauthorblockA{Department of Electrical and Computer Engineering,\\ Iowa State University, Ames, Iowa, 50011\\
Email: \{atamrawi,kothari\}@iastate.edu}}


%Telephone: (800) 555--1212, Fax: (888) 555--1212}


\maketitle

\begin{abstract}
Efficient and accurate path-sensitive analyses pose the challenges of: (a) analyzing an exponentially-increasing number of paths in a control-flow graph (CFG), and (b) checking feasibility of paths in a CFG. We address these challenges by introducing an \emph{equivalence relation} on the CFG paths to partition them into equivalence classes. It is then sufficient to perform analysis on these equivalence classes rather than on the individual paths in a CFG. This technique has two major advantages: (a) although the number of paths in a CFG can be exponentially large, the essential information to be analyzed is captured by a small number of equivalence classes, and (b) checking path feasibility becomes simpler. The key challenge is \emph{how to efficiently compute equivalence classes of paths in a CFG without examining each path in the CFG?} In this paper, we present a \emph{linear-time} algorithm to form equivalence classes without the need for examination of each path in a CFG. The key to this algorithm is construction of an \emph{event-flow graph} (EFG), a compact derivative of the CFG, in which each path represents an equivalence class of paths in the corresponding CFG. EFGs are defined with respect to the set of events that are in turn defined by the analyzed property. The equivalence classes are thus guaranteed to preserve all the event traces in the original CFG. We present an empirical evaluation of the Linux kernel (v3.12). The EFGs in our evaluation are defined with respect to events of the spin safe-synchronization property. Evaluation results show that there are many fewer EFG-based equivalence classes compared to the corresponding number of paths in a CFG. This reduction is close to 99\% for CFGs with a large number of paths. Moreover, our controlled experiment results show that EFGs are human comprehensible and compact compared to their corresponding CFGs.
\end{abstract}
%\begin{keywords}
%D.2 [Software Engineering]: Software/Program Verification; D.2.10.c [Software Engineering]: Representation; D.2.17.b [Software Engineering]: Code Design;
%\end{keywords}
%\category{D.2}{Software Engineering}: {Software/Program Verification}
%\category{D.2.10.c}{Software Engineering}{Representation}
%\category{D.2.17.b}{Software Engineering}{Code Design}
%\terms
%Algorithms, Path-Sensitive

%\keywords
%Program analysis, Critical section, Concurrent programs, Static analysis

\section{Introduction}
\label{introduction}
Accurate path-sensitive analyses require that: (a) execution effects along each path in a control-flow graph (CFG) are analyzed in isolation, i.e., without merging effects from different paths, and (b) effects along infeasible paths are excluded. The specific challenges for efficient and accurate path-sensitive analyses are thus: (a) exponential growth of the number of paths with the number of non-nested branch nodes in the CFG \cite{das2002esp, carter2003folklore}, and (b) checking path feasibility requires checking satisfiability of branch conditions along the path, a process that also can incur exponential computation~\cite{ngo2007detecting, vojdani2009goblint, navabi2010path, ball2001bebop, xu2008path, dillig2008sound}.

Intuitively, our novel approach to efficient path-sensitive analysis works by considering equivalence classes of paths as follows: \emph{Two paths are considered equivalent if they have the same event trace}. Each event trace is a sequence of events on a CFG path representing sufficient information for checking whether a given software property holds on that path. Since all paths in an equivalence class have the same event trace, it is sufficient to check just one path per group/class. In theory, there is an opportunity to circumvent exponential computational growth if the number of equivalence classes remains small even as the number of CFG paths grows exponentially. Although it may seem counterintuitive to just compute the event trace for each class without explicitly examining each CFG path, in this paper, we present an innovative linear-time algorithm that computes all event traces without examining each CFG path, thereby circumventing the exponential computational load.

We define a derivative of CFG, called the \emph{Event-Flow Graph} (EFG), and prove that there is a one-to-one and onto mapping, where each path in the EFG corresponds to a group/class of equivalent paths in the CFG. The EFG retains the events and relevant branch nodes from the CFG, and the EFG is a minimal graph for computing all the event traces, i.e., each path in the EFG produces a unique event trace. We provide a linear-time algorithm to compute the EFG. Another benefit of introducing EFG is that it can minimize computation for checking path feasibility. In our empirical evaluation of the Linux kernel (v3.12) (Section~\ref{manual}), we found that only the relevant branch nodes for forming equivalence classes were also \emph{sufficient} for checking feasibility.
%Another potential benefit of introducing EFG is that it \emph{minimizes} computation for checking path feasibility as follows: the EFG retains only relevant branch nodes and it is sufficient to check the satisfiability of the relevant branch conditions on the corresponding path in the EFG. The correlation between the relevant conditions can be inferred via constant prorogation~\cite{wegman1991constant} or global value numbering~\cite{click1995global}.
%Another potential benefit of introducing EFG is that it \emph{minimizes} computation for checking path feasibility as the EFG retains only relevant branch nodes. A particular event trace can occur on many CFG paths and each CFG path can have many branch nodes, so instead of needing to check that all branch conditions are satisfied on each of these CFG paths, it is sufficient to check that relevant branch conditions are satisfied on the corresponding path in the EFG. The correlation between the relevant conditions can be inferred via constant prorogation~\cite{wegman1991constant} or global value numbering~\cite{click1995global}.

To assess the benefits of using EFGs in a practical scenario, we conducted an empirical evaluation of the Linux kernel (v3.12). The EFGs are defined with respect to events relevant for verifying the \emph{spin safe-synchronization property} that requires that a lock of a spin object is followed by an unlock of the same object on all feasible execution paths. Our results show that the number of paths and the number of branch nodes are drastically reduced in going from a CFG to its EFG. The results from our controlled experiment show that EFGs are human comprehensible and compact compared to their corresponding CFGs and that is apparent in the reduction of the manual verification time and the effort in checking feasibility. The controlled experiment resulted on reporting a new bug~\cite{LinuxBug} that has been accepted by the Linux community. All the CFGs and their corresponding EFGs from our empirical evaluation are available in~\cite{RESULTS}. In Section~\ref{case-study}, we present an example using the Linux kernel to illustrate the use of EFG in an intra- and inter-procedural path-sensitive analysis.

%The remainder of the paper is organized as follows. Section 2 presents a motivating example. Next, Section 3 introduces the notions of event traces and irrelevant branch nodes. It also defines the equivalence relation of the CFG paths. Section 4 describes a linear-time algorithm for computing equivalence classes of paths in CFGs. Section 5 presents our empirical evaluation. Section 6 shows a case study of using EFGs in path-sensitive analysis. Related work and conclusions are presented in Sections 7 and 8 respectively.

\section{A Motivating Example}
\label{motivation}

\begin{figure}[ht]
    \centering
    \includegraphics[width = 0.35\textwidth]{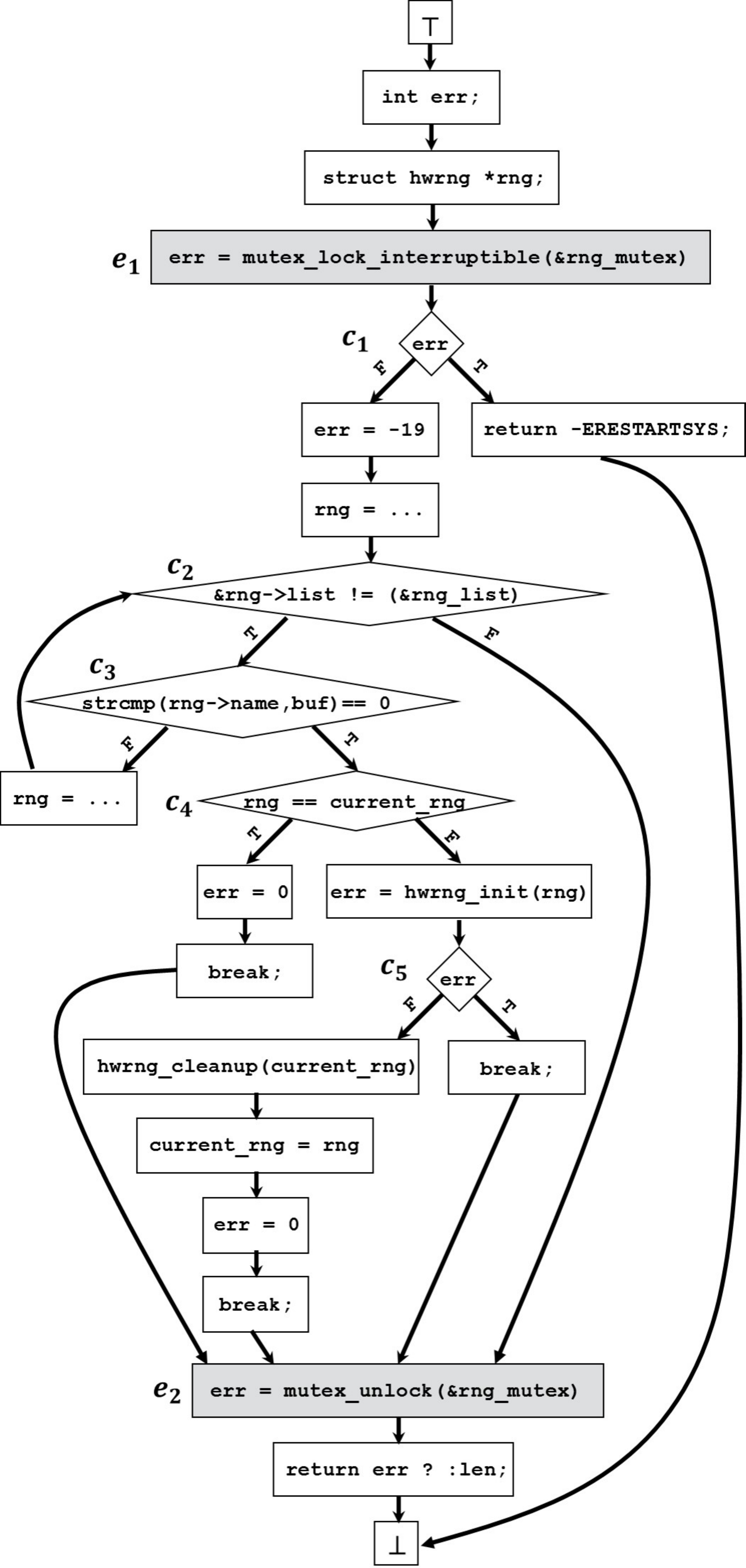}
    \caption{CFG for Function \code{hwrng\_attr\_current\_store}}
    \label{fig:motivation-cfg}
    \vspace{-16pt}
\end{figure}

Figure~\ref{fig:motivation-cfg} shows the CFG \remove{control flow graph (CFG)} of function \code{hwrng\_attr\_current\_store} from the Linux kernel (v3.12). The nodes $\top$ and $\bot$ respectively denote the unique entry and exit nodes added to the CFG. Also, \code{T} and \code{F} respectively denote the \code{true} and \code{false} branches from a branch node. This example concerns event traces that are needed to check the \emph{safe-synchronization property} that requires that a lock of a synchronization object must be followed by an unlock of the same locked object over all feasible CFG paths. The example contains two event nodes highlighted in gray: \code{mutex\_lock\_interruptible (\&rng\_mutex)} ($e_1$) and \code{mutex\_unlock (\&rng\_mutex)} ($e_2$).

In the CFG, the branch nodes $c_2, c_3, c_4,$ and $c_5$ are considered \emph{irrelevant branch nodes}; a branch node is \emph{irrelevant} if all paths branching from it lead to the same \emph{event} or \emph{terminal} node. The branch node $c_1$ is a \emph{relevant} branch node. For a given path $P$ in a CFG, the sub-trace of the execution trace of $P$, consisting of only the \emph{relevant branch} and \emph{event} nodes, is called the \emph{event trace} of $P$. That means, the paths branching from the \code{false} branch of $c_1$ have the same \emph{event trace} ($\top e_1 c_1 e_2 \bot$). Paths with identical event traces are considered \emph{equivalent} and grouped into one \emph{equivalence class}. The event trace of the path branching from the \code{true} branch of $c_1$ is $\top e_1 c_1 \bot$, and it is by itself in another equivalence class.

Each equivalence class is represented by a unique event trace of the paths in that class. Once those equivalence classes are computed, it is sufficient to analyze \emph{only} the event traces for all equivalence classes to check the safe-synchronization property. For example, the CFG in Figure~\ref{fig:motivation-cfg} results in two event traces: 1) the trace $\top e_1 c_1 e_2\bot$, and 2) the trace $\top e_1 c_1 \bot$. It is then sufficient to analyze only these two traces to cover all the paths in the CFG.

To summarize the important points:
\begin{enumerate}
    \item All CFG paths with the same event trace are grouped into one equivalence class. In this example, all the CFG paths are grouped into two equivalence classes.
    \item Analyzing event traces is equivalent to analyzing all CFG paths.
    \item A branch node is \emph{irrelevant} if all the paths branching from it either lead to the same event or terminal node (a broader notion of irrelevant branch nodes is defined later). Four out of five branch nodes are irrelevant in this example.
\end{enumerate}
\section{Path-Sensitive Analysis with Event Traces}
\label{terminology}

\begin{mydef}
A \textbf{Control Flow Graph} (CFG) of a program is defined as $G = (V, E, \top, \bot)$, where $G$ is a directed graph with a set of nodes $V$ representing the program statements and a set of edges $E$ representing the control-flow between statements. $\top$ and $\bot$ denote the respective unique entry and exit nodes of the graph.
\end{mydef}

\subsection{Execution Traces}
\begin{figure}[ht]
    \centering
    \includegraphics[width = 0.43\textwidth]{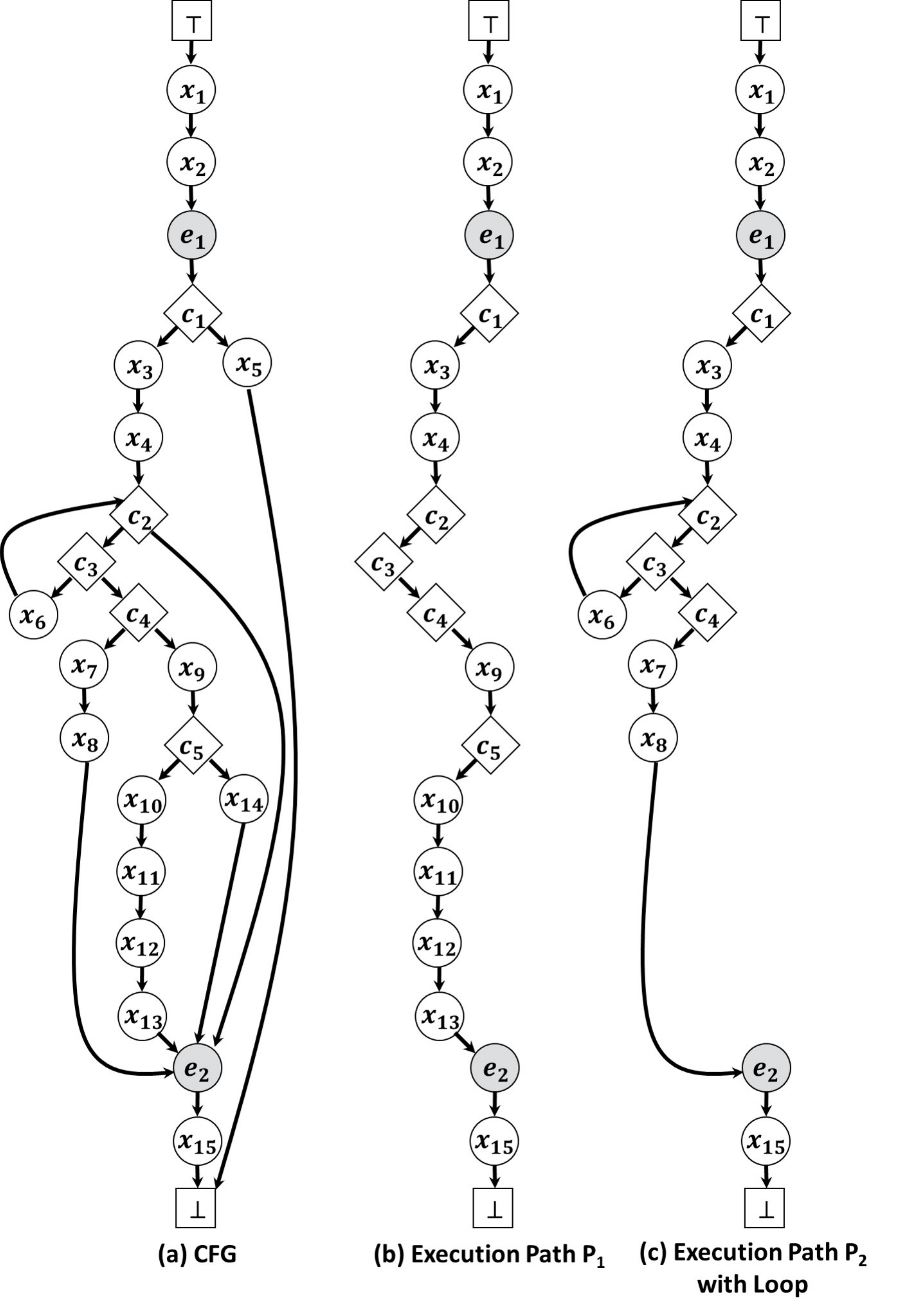}
    \caption{Execution paths in a CFG}
    \label{fig:compact-cfg}
\end{figure}

After labeling each node of the CFG with a unique identifier, we define the \emph{execution trace} of a CFG path as the regular expression of the node labels along that path. Let us illustrate this definition using the CFG first shown in Figure~\ref{fig:motivation-cfg} and redrawn in Figure~\ref{fig:compact-cfg}(a). The highlighted nodes are the event nodes. Code statements are replaced with nodes labeled $x_1$ through $x_{15}$. Figure~\ref{fig:compact-cfg}(b) shows a CFG  path; its \emph{execution trace} is: $\top x_1 x_2 e_1 c_1 x_3 x_4 c_2 c_3 c_4 x_9 c_5 x_{10} x_{11} x_{12} x_{13} e_2 x_{15} \bot$. Figure~\ref{fig:compact-cfg}(c) shows a CFG path with a loop; its \emph{execution trace} is: $\top x_1 x_2 e_1 c_1 x_3 x_4 (c_2 c_3 x_6)^{+} c_2 c_3 c_4 x_7 x_8 e_2 x_{15} \bot$, where $(n_i .. n_k)^{+}$ represents the loop between the nodes $n_i$ to $n_k$ that is executed one or more times.

\subsection{Event Traces}
Event traces are defined with respect to $\mathcal{E}$; the set of events of interest associated with the analyzed property. We will use the verification of the \emph{safe-synchronization property} as a running example in illustrating all definitions. The safe-synchronization property requires that for every object $p$: the locking event $e_1(p)$ is followed by the unlocking event $e_2(p)$ on all feasible execution paths. Thus, the set $\mathcal{E}_p$ of events, for verifying that property for object $p$, consists of: the \emph{locking} and \emph{unlocking} events defined on $p$, and the \emph{data-flow} events in which the locked object $p$ is either aliased or escapes to another function as a parameter, a return value, or a global variable. Such data-flow events can be determined using a taint analysis technique~\cite{ceara2009detecting}. Once the events of interest are determined, one can write a path-sensitive verification algorithm that traverses the CFGs and checks that, for each synchronization object ($p$), a \emph{lock} event $e_1$($p$) is always succeeded by an \emph{unlock} event $e_2$($p$) on all \remove{feasible} execution paths. In case of a \emph{violating path} which has a lock event not followed by unlock event, path feasibility needs to be conducted.

\begin{mydef}
A property $P$ is a \emph{2-event property} if for every object $p$, an event $e_1(p)$ must be succeeded by another event $e_2(p)$ on all feasible execution paths.
\end{mydef}

Note that event-based analyses can be performed to verify properties that can be modeled as $2$-event properties like safe-synchronization and memory leak. A number of vulnerabilities listed by the MITRE Corporation~\cite{CWE} can be addressed using event-based analyses.

\begin{mydef}
A CFG node is an \emph{event} node if it corresponds to an event of the set of events ($\mathcal{E}$) associated with a given property to be analyzed.
\end{mydef}

\begin{mydef}
Successors of a node $u$ in a directed graph $G$, denoted by \emph{suc($u$)}, consist of the set of nodes $v \neq u$  such that $\exists$ an edge $(u, v)$.
\end{mydef}

\begin{mydef}
Successors of a subgraph $S$ in a directed graph $G$, denoted by \emph{suc($S$)}, consist of the set of nodes $v \notin S$ such that $v = $\emph{suc($u$)} for $u \in S$.
\end{mydef}

\begin{mydef}
For a branch node $c$, a \textbf{branch edge} is an out-coming edge of $c$.
\end{mydef}

\begin{mydef}
\label{irrelevant-branch-nodes}
A branch node $c$ is an \textbf{irrelevant branch node} if the following conditions are satisfied:
\begin{itemize}
    \item $c$ is a non-event node.
    \item There exists a subgraph $S$ containing $c$ and all branch edges of $c$.
    \item $S$ has no event nodes.
    \item $S$ has a unique successor, i.e., $|$\emph{suc($S$)}$|=1$.
\end{itemize}
\end{mydef}

\begin{figure}[ht]
    \vspace{-8pt}
    \centering
    \includegraphics[width = 0.23\textwidth]{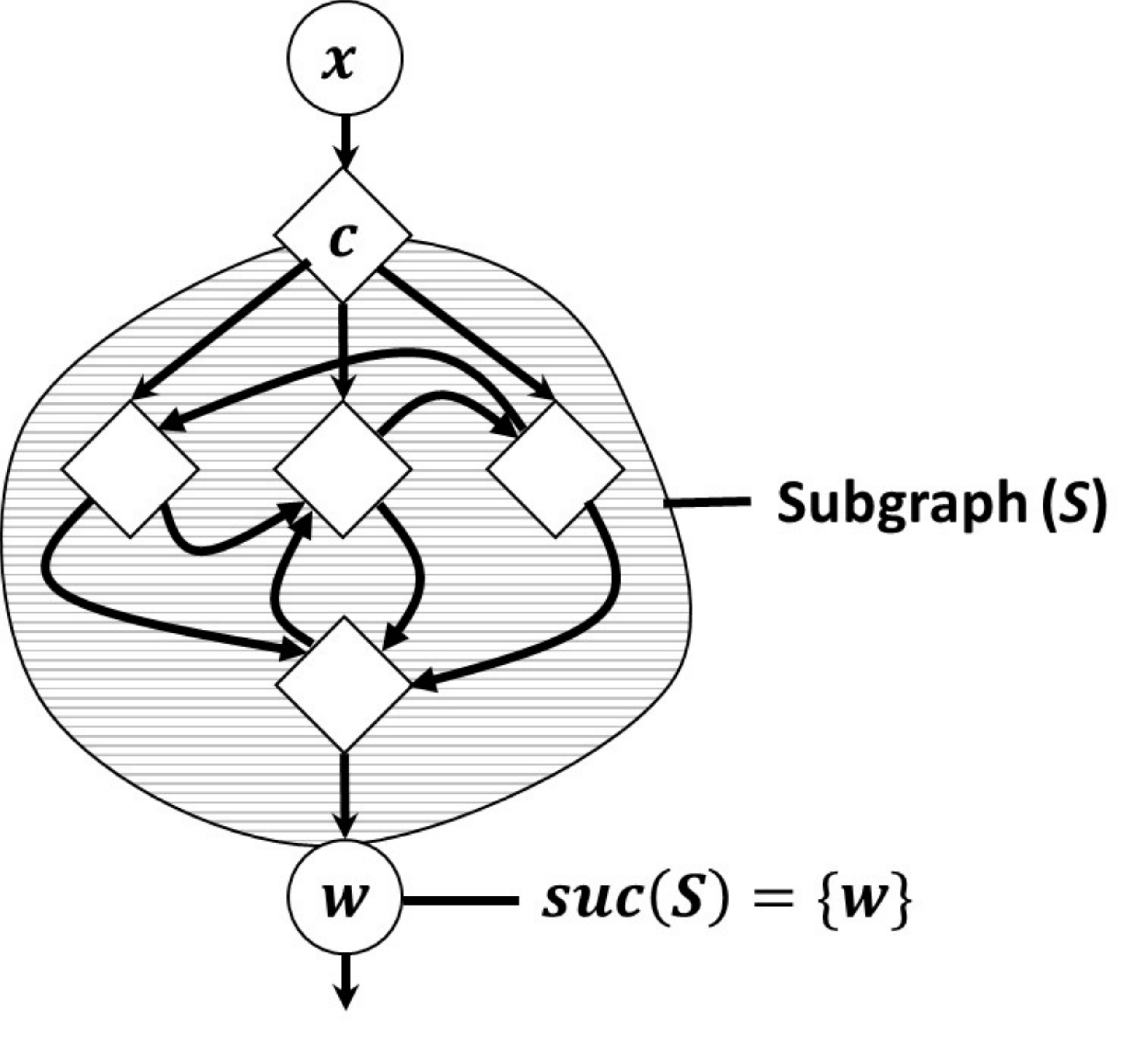}
    \caption{An example of irrelevant branch node $c$}
    \label{fig:irrelevant-branch-node}
    \vspace{-2pt}
\end{figure}

\noindent Figure~\ref{fig:irrelevant-branch-node} shows an example of irrelevant branch node $c$.

\begin{mydef}
An \emph{event trace} of a path in CFG is a \emph{sub-expression} of the execution trace consisting only of the \emph{relevant branch} nodes, \emph{event} nodes and the $\top$ and $\bot$ nodes.
\end{mydef}

For example, the event trace for paths $P_1$ and $P_2$ depicted in Figure~\ref{fig:compact-cfg}(b, c) would be: $\top e_1 c_1 e_2 \bot$.

\begin{mydef}
The \textbf{Event-Flow Graph} (EFG) $G_{\text{EFG}}$ of a CFG $G$ with respect to $\mathcal{E}$ is the node-induced subgraph of $G$ consisting of the event nodes, the relevant branch nodes, and the entry ($\top$) and exit ($\bot$) nodes.
\end{mydef}

%\noindent \textbf{Relevancy of Branch Nodes.} Our notion of relevancy is defined with respect to forming equivalence classes of CFG paths. For checking feasibility, the notion of relevancy would be different in theory. That means, the correlation between relevant branch nodes may depend on other irrelevant branch nodes or non-event nodes. However, our empirical evaluation of the Linux kernel (Section~\ref{manual}) shows that the relevant branch nodes are sufficient for checking feasibility.

\noindent \textbf{Relevancy of Branch Nodes.} The notion of relevancy is defined with respect to equivalence classes of CFG paths. There could be \emph{correlation} between relevant branch nodes and the remaining branch nodes or non-event nodes. In that case, the path feasibility check would need to take into account other nodes correlated to relevant branch nodes. In our empirical evaluation of the Linux kernel (Section~\ref{manual}), we did not find such correlation. Thus, only the relevant branch nodes for forming equivalence classes were also \emph{sufficient} for checking feasibility.

\subsection{Equivalence Classes of CFG Paths}
\label{problem}
In this section, we will show that verifying a property $P$ on all CFG paths \emph{is equivalent} to verifying $P$ for all event traces. Initially, let us define an \emph{equivalence relation} on the CFG paths:
\begin{mydef}
Given a CFG $G$ and an event set $\mathcal{E}$ which is a subset of nodes in $G$, we define a relation $\mathcal{R}_\mathcal{E}$ on paths in $G$ as follows: two CFG paths are \textbf{related} iff they have the same event trace.
\end{mydef}

Note that $\mathcal{R}_\mathcal{E}$ is an equivalence relation because it is \emph{reflexive}, \emph{symmetric}, and \emph{transitive}.

\begin{theorem}\footnote{Theorems' \& corollaries' proofs can be found in appendix~\ref{appendix}.}
\label{theorem-1}
Given a CFG $G$, a set $\mathcal{E}$ of events, and the equivalence relation $\mathcal{R}_\mathcal{E}$, there is a one-to-one and onto mapping between the equivalence classes of $\mathcal{R}_\mathcal{E}$ and the paths of the EFG $G_{\text{EFG}}$ where each EFG path produces the event trace corresponding to an equivalence class.
\end{theorem}
\remove{\noindent\textbf{Proof.} The equivalence relation $\mathcal{R}_\mathcal{E}$ partitions the CFG paths into equivalence classes such that all paths in an equivalence class have the same event trace, and the CFG paths that are in different equivalence classes have different event traces.

Since $G_{\text{EFG}}$ is the node-induced subgraph of the given CFG $G$ consisting of the event and the relevant branch nodes, it follows that given an EFG path $S$, it will produce an unique event trace $T$ and conversely given an event trace $T$ there will be a unique EFG path $S$ for which the event trace is $T$. Thus, there is a one-to-one and onto mapping between the equivalence classes of $\mathcal{R}_\mathcal{E}$ and the paths of the EFG $G_{\text{EFG}}$.}

\begin{theorem}
\label{theorem-2}
Given a \emph{2-event property} $P$, its verification on all CFG paths can be done using the event traces.
\end{theorem}

\begin{corollary}
Given a \emph{2-event property} $P$, its verification on all paths of a CFG $G$ can be done with the corresponding EFG $G_{\text{EFG}}$.
\end{corollary}

\noindent The proof directly follows from the above two theorems.

\begin{remark}
Given a CFG and the set $\mathcal{E}$, the corresponding EFG is unique. This is because the nodes of EFG are uniquely defined as they are exactly the event nodes and relevant branch nodes. The edges in the EFG are also uniquely defined because they are induced by the edges in the CFG.
\end{remark}

\subsection{EFGs for Optimal Path-Sensitive Analyses}
\label{steps}
Given a \emph{2-event property}, its path-sensitive analysis can be done using EFGs instead of CFGs as follows:

\noindent For every object $p$:
\begin{enumerate}
    \setlength{\itemsep}{0pt}
    \setlength{\parskip}{0pt}
    \setlength{\parsep}{0pt}
\item Determine all the events of interest (i.e., the set $\mathcal{E}_p$), including the events $e_1$ and $e_2$ defined on $p$ and the data-flow events for $p$.
\item If the object $p$ is passed to other functions either as a parameter or as a return value, then consider all such functions to be \emph{relevant} functions.
\item Construct EFGs with respect to $\mathcal{E}_p$ for \emph{relevant} functions.
\item Perform the path-sensitive analysis using the EFGs instead of the CFGs.
\item In case of a path that has $e_1(p)$ not followed by $e_2(p)$, path feasibility check is needed. Using EFGs, the path feasibility is conducted by checking the satisfiability of the conditions (relevant branch nodes), where the correlation between conditions can be computed via constant propagation~\cite{wegman1991constant} or global value numbering~\cite{click1995global} as in~\cite{cho2013}.
\end{enumerate}

EFGs can be used to perform inter-procedural path-sensitive analysis. In Section~\ref{case-study}, we illustrate a case study on inter-procedural verification using EFGs. To conclude, EFG is the minimal graph that produces all event traces for an accurate and efficient path sensitive analysis given a \emph{2-event property} $P$ and its associated set of events $\mathcal{E}$. For the EFG to be useful in practice, the next section presents an efficient algorithm to construct the EFG from a given CFG.
%For example, the event $e_1(p)$ occurs in function $f_1$, then object $p$ is passed as a parameter $q$ to another function $f_2$ that is called on a CFG path in $f_1$, and the event $e_2(q)$ occurs in function $f_2$. Let $E_1$ and $E_2$ denote the EFGs of $f_1$ and $f_2$ respectively. $E_2$ is linked to $E_1$ at the node where $f_2$ is called in $f_1$. At this call-site node in $E_1$, the analysis enters and traverses through $E_2$ and, at the exit from $E_2$, returns to $E_1$ to continue the traversal in $E_1$. The linkage between $E_1$ and $E_2$ can be conveniently established because of the unique entry ($\top$) and exit ($\bot$) nodes. An example of inter-procedural path-sensitive verification is provided later in an application study given in Section~\ref{case-study}.

%So far, we have defined the EFG as a node-induced subgraph consisting of the event nodes and the relevant branch nodes. Each path in the EFG represents a unique event trace and every event trace is given by some path in the EFG. \remove{We have shown that a path-sensitive verification of a \emph{2-event property} $P$ on all CFG paths can be done with the corresponding EFG.} Thus, the EFG is the minimal graph that produces all event traces for an accurate and efficient path sensitive analysis given a \emph{2-event property} $P$ and its associated set of events $\mathcal{E}$. For the EFG to be useful in practice, \remove{we need an efficient algorithm to construct the EFG from a given CFG.} the next section presents an efficient algorithm to construct the EFG from a given CFG. 
\section{Computing Event Traces of a CFG}
\label{approach}
In this section, we first present an algorithm to compact a given CFG into a \emph{T-irreducible} graph. This compaction algorithm will be used later in computing the EFG. Then, we present an algorithm that \emph{efficiently} computes the equivalences classes of CFG paths in linear-time based on Tarjan's algorithm to compute strongly-connected components of a directed graph~\cite{tarjan1972depth}.

\subsection{Algorithm I: T-irreducible Graph}
\label{algo1}
\begin{mydef}
A \textbf{Colored Directed Graph} (CDG) is defined as $G = (V, E, C, \top, \bot)$, where $(V, E)$ is a finite directed graph with a set of nodes $V$ and a set of edges $E$. $\top$ and $\bot$ respectively represent the unique entry and exit nodes of the graph. $C \subseteq V$ is the set of colored nodes.
\end{mydef}
\noindent For the purpose of this algorithm, a CFG is modeled as a CDG by treating the event nodes as colored nodes.\vspace{2pt}

\noindent \textbf{$T_1$: Elimination of Non-branching and Non-colored Nodes}

\begin{figure}[ht]
    \centering
    \includegraphics[width = 0.48\textwidth]{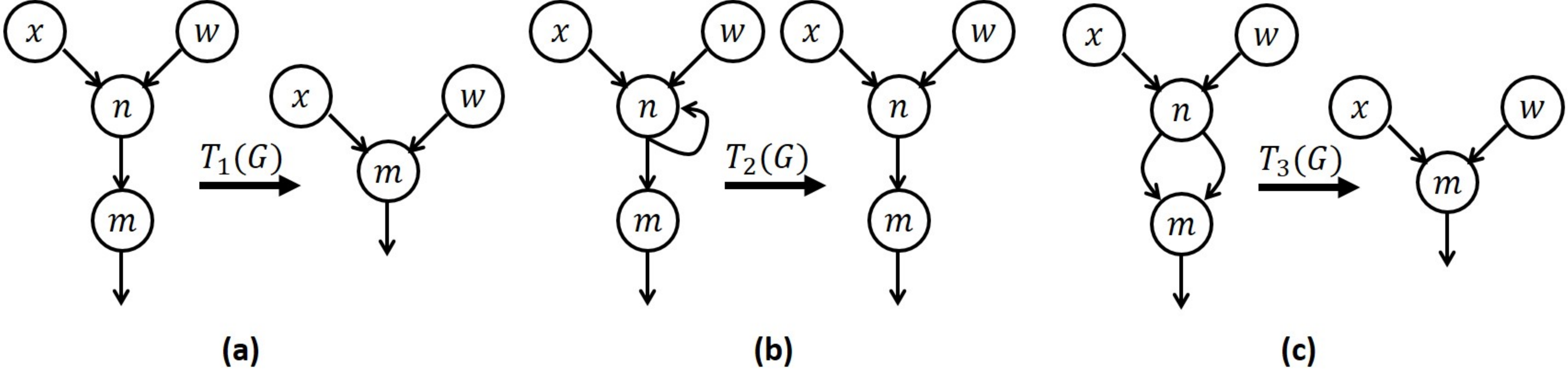}
    \caption{Algorithm I transformations: (a)$T_1$, (b)$T_2$, (c)$T_3$}
    \label{fig:algorithm2-transformations}
    \vspace{-8pt}
\end{figure}

\noindent Let $G = (V, E, C, \top, \bot)$ be a CDG and $n$ be a \emph{non-colored} node ($n \notin C$) with a single successor $m$. The $T_1$ transformation is the consumption of node $n$ by $m$. Induced edges are introduced so that the predecessors of node $n$ become predecessors of node $m$. (Figure~\ref{fig:algorithm2-transformations}(a))
% and, in particular, an edge $(m, n)$ becomes an edge $(m, m)$.

%If $P_n$ is the set of predecessor nodes of $n$ then the resulting CDG $G^\prime = T_1 (G) = (V - \{n\}, (E - \{(i, n) \quad i \in P_n\}) \bigcup \{(i, m) \quad i \in P_n\}, C, \top, \bot)$. In short $G \overset{T_1 (G)}{\Rightarrow} G^\prime$. (Figure~\ref{fig:algorithm2-transformations}(a))

The $T_1$ transformation eliminates every node from CFG that is neither a branch node nor an event node. These nodes are removed because they are irrelevant to the analysis because they are \textbf{not} included in the resultant event traces.

\vspace{5pt} \noindent \textbf{$T_2$: Elimination of Self-Loop Edges}

\noindent Let $G = (V, E, C, \top, \bot)$ be a CDG and let $n$ be a \emph{non-colored} node ($n \notin C$) that has a self-loop edge $(n, n)$. The $T_2$ transformation removes that edge. (Figure~\ref{fig:algorithm2-transformations}(b))

%Then, the resulting CDG $G^\prime = T_2 (G) = (V, E -\{(n, n)\}, C, \top, \bot)$. In short $G \overset{T_2 (G)}{\Rightarrow} G^\prime$. (Figure~\ref{fig:algorithm2-transformations}(b))

The intuition behind $T_2$ transformation is: in a loop block that contains no event nodes, execution of the loop is immaterial. Therefore, $T_2$ removes the self-loop edges.% $T_2$ also removes the self-loop edges resulting from the $T_1$ transformation.

\vspace{5pt} \noindent \textbf{$T_3$: Elimination of Irrelevant Branch Nodes}

\noindent Let $G = (V, E, C, \top, \bot)$ be a CDG and let $n$ be a \emph{non-colored} node ($n \notin C$) that has two or more edges, all pointing to the same successor $m$ of $n$. Then the $T_3$ transformation is the consumption of node $n$ by $m$ and the predecessors of node $n$ become predecessors  of node $m$. (Figure~\ref{fig:algorithm2-transformations}(c))
%In particular, an edge $(m, n)$ becomes an edge $(m, m)$.

%If $P_n$ is the set of predecessor nodes of $n$ then the resulting CDG $G^\prime = T_3 (G) = (V - \{n\}, (E - \{(i, n) \quad i \in P_n\}) \bigcup \{(i, m) \quad i \in P_n\}, C, \top, \bot)$. In short $G \overset{T_3 (G)}{\Rightarrow} G^\prime$. (Figure~\ref{fig:algorithm2-transformations}(c))

The intuition behind the $T_3$ transformation is as follows: Imagine the case where a branch node $n$ has only non-colored nodes on its branches, and all those branches ultimately merge at node $m$. If the non-colored nodes on those branches are eliminated by the $T_1$ transformation, all branches will point to node $m$. At this point, the branching at $n$ is \emph{irrelevant} so the branch node $n$ can be eliminated.

\begin{mydef}
Let $T = \{T_1, T_2, T_3\}$ be the set of basic transformations described above. A CDG $G$ is \emph{$T$-irreducible} if it cannot be further reduced by applying transformations in $T$.
\end{mydef}

Algorithm I uses the transformations $T_1$, $T_2$, and $T_3$ to construct a \emph{T-irreducible} graph as follows:
\begin{itemize}
    \item Start with a CFG $G_{\text{CFG}} = (V_{G_{\text{CFG}}}, E_{G_{\text{CFG}}}, \top, \bot)$ and transform it into a CDG $G_{\text{CDG}} = (V_{G_{\text{CFG}}}, E_{G_{\text{CFG}}}, C, \top, \bot)$. The set $C$ of colored nodes is the set of event nodes defined by the set $\mathcal{E}$ associated with the analyzed property. i.e., CFG is modeled as a CDG by treating the event nodes as colored nodes.
    \item Transform $G_{\text{CDG}}$ into a \emph{T-irreducible} graph by applying the transformations in $T$.
\end{itemize}

\subsection{Algorithm II: Transform CFG to EFG}
\label{algo2}
We will present a \emph{linear-time} algorithm that produces the minimal number of equivalence classes by transforming a CFG $G$ into $G_{\text{EFG}}$, the corresponding \emph{event-flow graph} (EFG). Intuitively, one might think that the problem of computing all event traces would require individual examination of each path in a CFG, but Algorithm II shows this not to be true. As observed in our empirical evaluation of the Linux kernel, while the number of CFG paths may grow exponentially, the number of event traces does not. In such scenarios, Algorithm II is very efficient because it requires a computational load proportional to the number of event traces rather than to the number of CFG paths.

We claim that the graph produced by Algorithm II is indeed the \emph{event-flow graph} (EFG) that we have defined earlier and shown to be the minimal graph that produces all the event traces needed for an accurate and efficient path-sensitive analysis. First, we will present CFG to EFG transformation algorithm and later provide a proof that the produced graph is actually the EFG. We will use the following additional definition to further describe Algorithm II.

\begin{mydef}
$G_{\text{CG}}$ is the \emph{condensation graph} of a directed graph $G$ if each \emph{strongly-connected component} (SCC) of $G$ \emph{contracts} to a single node in $G_{\text{CG}}$ and the edges of $G_{\text{CG}}$ are induced by edges in $G$. Thus, $G_{\text{CG}}$ is a directed acyclic graph (DAG).
\end{mydef}

%\remove{It is shown in~\cite{SCC} that the \emph{condensation graph} is a directed acyclic graph (DAG). }
Given a CFG $G_{\text{CFG}}$ and the set $\mathcal{E}$ of events as computed in step (1) in Section~\ref{steps}. Now, let us fully describe algorithm~II that transforms the CFG $G_{\text{CFG}}$ to its EFG $G_{\text{EFG}}$:

\noindent \textbf{(1) T-Irreducible Graph Construction:} Start with a CFG $G_{\text{CFG}}$ and transform it into a \emph{T-irreducible} graph $G_{\text{T-irr}}$ by applying Algorithm I.

\noindent \textbf{(2) Non-Colored Condensation Graph Construction:} Compute the subgraph $G_{\text{I}}$ of $G_{\text{T-irr}}$ induced by its non-colored nodes. Then, construct the non-colored condensation graph $G_{\text{NCCG}}$ of $G_{\text{I}}$.

\noindent \textbf{(3) Colored Condensation Graph:}  Construct a new CDG $G_{\text{CCG}}$ by adding the colored nodes in $G_{\text{T-irr}}$ to $G_{\text{NCCG}}$. If an edge exists between an SCC and a colored node $n$ in $G_{\text{T-irr}}$ then introduce an edge in $G_{\text{CCG}}$ between the contracted node for that SCC and the colored node $n$.

\noindent \textbf{(4) Condensed EFG Construction:} Transform $G_{\text{CCG}}$ into a \emph{$T$-irreducible} graph $G_{\text{cEFG}}$ by applying Algorithm I.

\noindent \textbf{(5) EFG Construction:} Transform $G_{\text{cEFG}}$ into $G_{\text{EFG}}$ by expanding each \emph{remaining} contracted SCC in $G_{\text{cEFG}}$ back to the original SCC as in $G_{\text{T-irr}}$.

\begin{remark}
The resultant graph $G_{\text{cEFG}}$ after step (4) is the \emph{condensed EFG}. In addition, we claim that the resultant graph $G_{\text{EFG}}$ after step (5) is the EFG.
\end{remark}

\begin{figure}[ht]
    \centering
    \includegraphics[width = 0.35\textwidth]{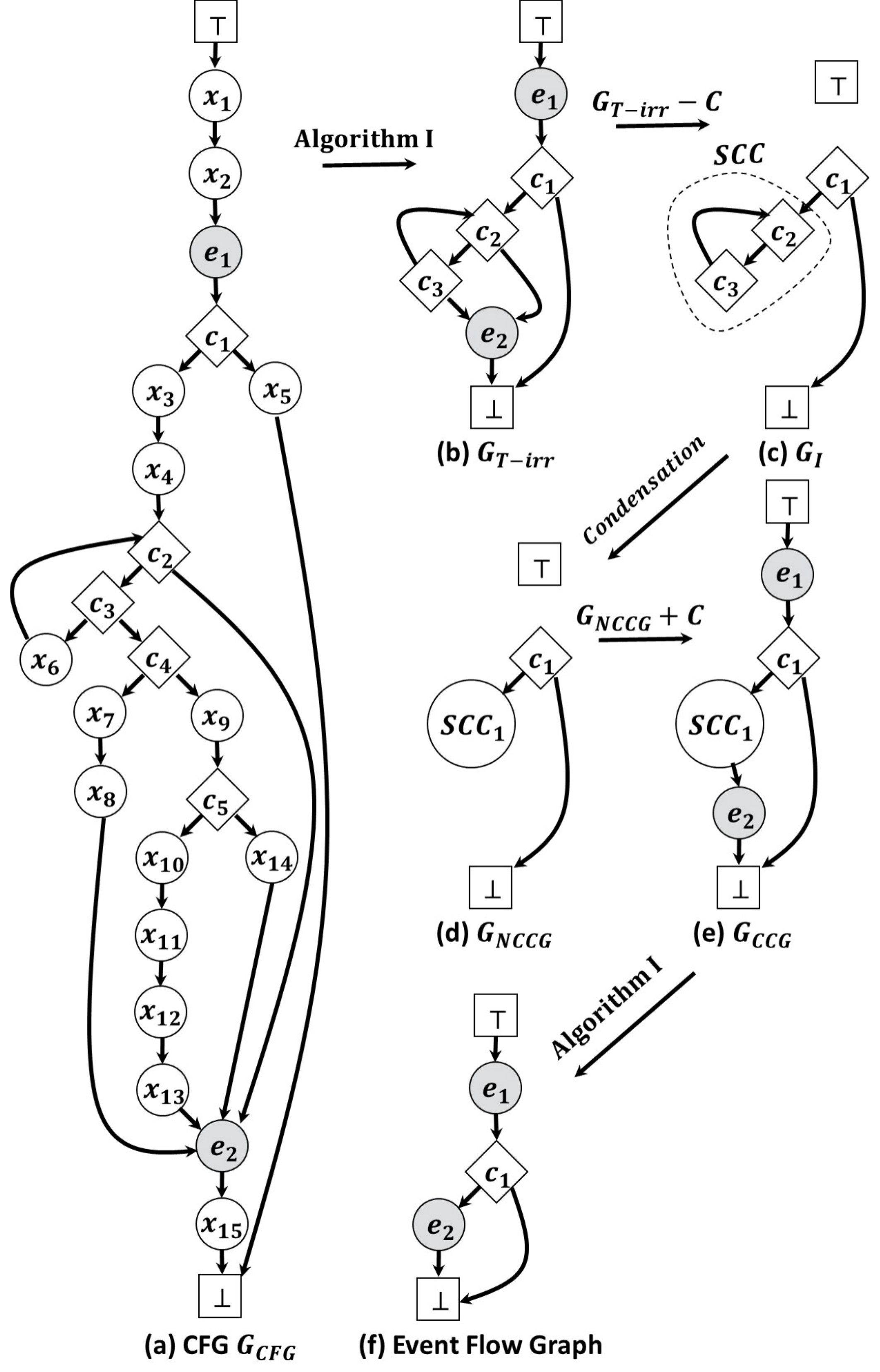}
    \caption{A transformation from CFG to EFG}
    \label{fig:example-transformation}
\end{figure}

\noindent Figures~\ref{fig:example-transformation}($a$-$f$) illustrate the successive graphs constructed by Algorithm II, starting with the CFG (graph $a$) and ending with the EFG (graph $f$).

\subsection{Algorithm II Complexity}
The algorithmic complexity of constructing the T-irreducible graph (Steps 1 and 4) is $O(|V| + |E|)$ where $|V|$ and $|E|$ are the respective numbers of nodes and edges in the CFG. For detecting the SCCs in step (2), we use an algorithm by Tarjan~\cite{tarjan1972depth} to compute strongly-connected components of a directed graph. The run-time of this algorithm is also $O(|V| + |E|)$, yielding a linear run-time complexity of $O(|V| + |E|)$ for Algorithm II.
%\subsection{Revisiting the Motivating Example}
%\label{revisit-motv}
%Figure~\ref{fig:example-transformation} shows the CFG to EFG transformation enacted by Algorithm III on the CFG of the function \code{hwrng\_attr\_current\_store} (Figure~\ref{fig:motivation-cfg}). The highlighted nodes are the event nodes and code statements at nodes are replaced with symbols $x_1$ through $x_{15}$. The figures $a$ through $f$ illustrate the successive graphs constructed by Algorithm III, starting with the CFG and ending with the EFG.
%\begin{figure}[ht]
%    \centering
%    \includegraphics[width = 0.35\textwidth]{new-figures/algorithm3-example-transformations}
%    \caption{A transformation from CFG to EFG}
%    \label{fig:example-transformation}
%\end{figure}
\subsection{Algorithm I versus Algorithm II}
The EFG constructed by Algorithm II achieves an important compaction of the CFG that is not possible in Algorithm~I. Recall that the graph $G_{\text{T-irr}}$ produced by Algorithm~I may contain irrelevant branch nodes. For example, consider two branch nodes $A$ and $B$ with $suc(A) = \{B,E\}$ and $suc(B) = \{A,E\}$, where $E$ is an event node. The branch nodes $A$ and $B$ are irrelevant and should be eliminated. The subgraph consisting of the two branch nodes $A$ and $B$ has a unique successor $E$ and thus, they will be eliminated in the condensed EFG $G_{\text{cEFG}}$. Note that in the above scenario even though there can be more that two branch nodes that are successors of each other, as long as they have the same event node as the successor, all such branch nodes are irrelevant and will be eliminated by Algorithm II. Thus, by including steps (2-5), Algorithm II can achieve compaction beyond that of Algorithm~I.

\begin{figure}[ht]
\vspace{-10pt}
    \centering
    \includegraphics[width = 0.3\textwidth]{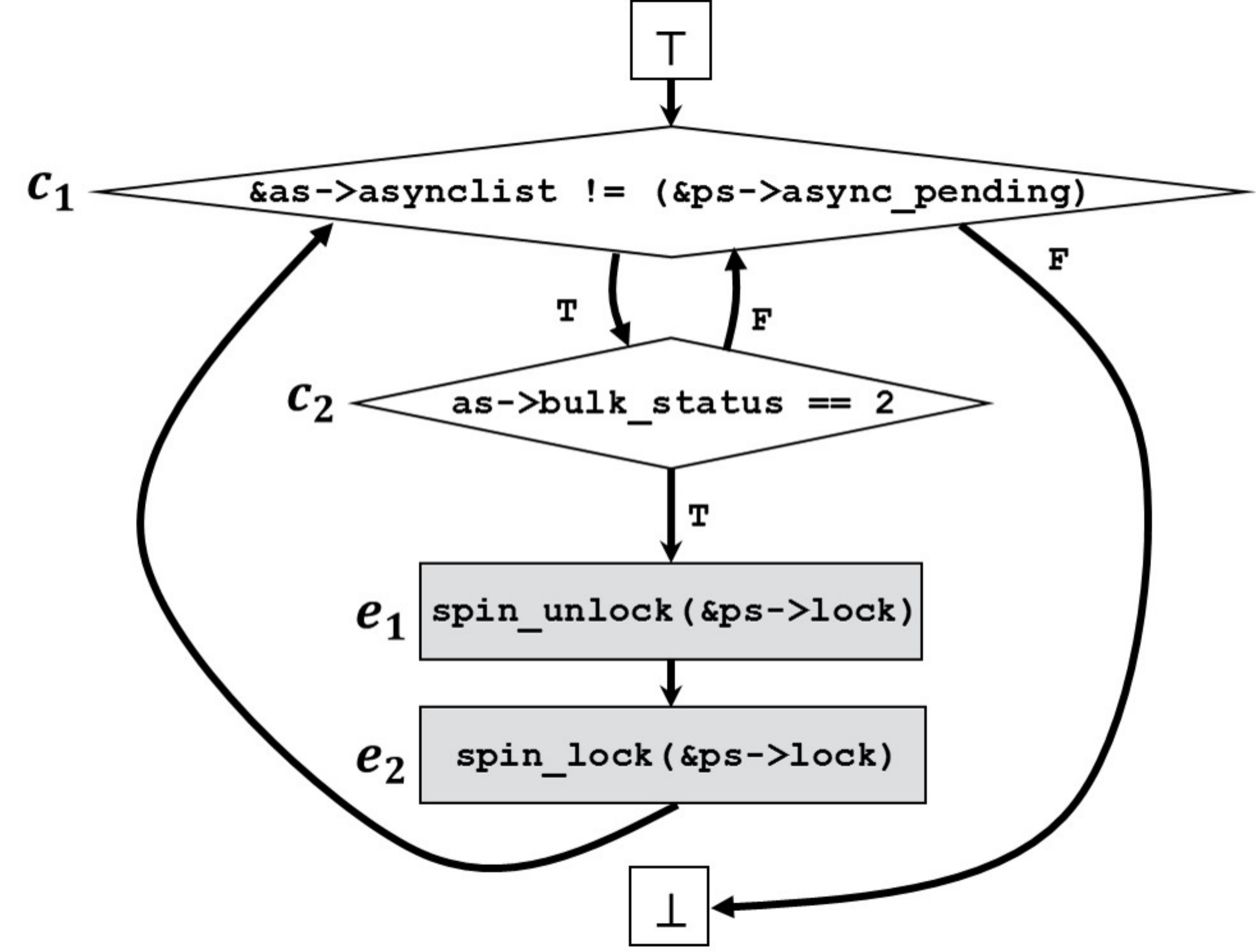}
    \caption{EFG for function \code{cancel\_bulk\_urbs} }
    \label{fig:efg-loop}
    \vspace{-10pt}
\end{figure}

Note that there can be a strongly-connected component (SCC) of branch nodes with two or more successors and, if so, those branch nodes must be retained. An example of such a scenario is shown in Figure~\ref{fig:efg-loop} depicting the EFG of function \code{cancel\_bulk\_urbs} from the Linux kernel (v3.12). In the EFG, the SCC consisting of the branch nodes $c_1$ and $c_2$ is retained as it has two successors: the terminal node $\bot$ and the event node $e_1$.
%Note that there can be a strongly-connected component (SCC) of branch nodes with two or more successors and, if so, those branch nodes must be retained. An example of such a scenario is shown in Figure~\ref{fig:efg-loop} depicting the EFG of function \code{cancel\_bulk\_urbs} from the Linux kernel (v3.12). The scenario involves a loop followed by an event ($e_1$), then followed by an event ($e_2$). The loop has two possible types of iterations; one does not involve any of the two events, and the second involves an event ($e_1$) followed by an event ($e_2$). The resulting behavior is summarized by the following event-trace expression $((c_1 c_2)^+ e_1 e_2)^+$ that is correctly captured by the EFG.

\subsection{EFG: Correctness Proof}
In Algorithm II, it is not enough to claim that the resultant graph $G_{\text{EFG}}$ after step (5) is the EFG. We will now prove that Algorithm II produces the EFG correctly. This amounts to proving that Algorithm II removes \emph{all} irrelevant branch nodes.

\begin{theorem}
\label{theorem-3}
Let $G$ be a colored \emph{$T$-irreducible} and acyclic graph. Then for any subgraph $S$ containing non-colored nodes of $G$: $|$\emph{suc($S$)}$| \geq 2$.
\end{theorem}

\begin{corollary}
\label{corollary-1}
\label{cefg}Let $G$ be a CFG and $G_{\text{cEFG}}$ be the condensed EFG. Then, for any subgraph $S$ containing non-colored nodes of $G_{\text{cEFG}}$, $|$\emph{suc($S$)}$|\geq 2$.
\end{corollary}

\begin{corollary}
\label{corollary-2}
The graph produced by Algorithm II does not contain any irrelevant branch nodes.
\end{corollary}

\section{An Empirical Evaluation}
We present an empirical evaluation to show the applicability and advantages of using event-flow graphs (EFG) rather than control-flow graphs (CFG) to perform path-sensitive analyses. Specifically, we evaluate the following:
\begin{itemize}
    \item The reduction in nodes and edges in going from CFGs to EFGs.
    \item The reduction in branch nodes, since only the \emph{relevant} branch nodes are retained in the EFGs.
    %\item A comparison between Algorithm II and Algorithm III.
    \item The reduction in the effort to check feasibility of paths.
\end{itemize}

\subsection{Experimental Setup}
We use the Linux kernel (v3.12) for our empirical evaluation. We consider the \emph{spin safe-synchronization property} for defining the events of interest ($\mathcal{E}$) for our analysis. We have used EnSoft's C-Atlas~\cite{ENSOFT} platform to compute $\mathcal{E}$, the CFGs, and EFGs. The C-Atlas first compiles the Linux kernel (v3.12) and pre-computes a \emph{database} of relationships (i.e., call, control-flow, data-flow, etc.) between various program artifacts. This process took 34 minutes. Then, we query the C-Atlas database to find all the \emph{spin synchronization} objects $\mathcal{P}$. This query identifies all the variables passed as parameters to the spin \emph{locking} (\code{spin\_lock, spin\_trylock}) and \emph{unlocking} (\code{spin\_unlock}) functions.

We followed the steps (1-3) in Section~\ref{steps} and wrote a Java program -based on a taint analysis technique~\cite{ceara2009detecting}- using the C-Atlas APIs to determine the set $\mathcal{E}_p$ for every object $p \in \mathcal{P}$, by identifying the events $e_1$ and $e_2$ that are defined on $p$ in addition to the relevant data-flow events for $p$, including those in which $p$ are passed as parameters or return values to other functions. Based on these events, we determine all the Linux kernel functions, referred to as \emph{relevant functions}, needing to be analyzed for every object $p$. Afterward, we wrote a Java program using the C-Atlas APIs to construct the EFG for each relevant function from its corresponding CFG based on the set $\mathcal{E}_p$.

In the Linux kernel (v3.12), there are 531 spin objects and in total 3,894 \emph{relevant} functions for the analysis of all objects. The conversion from CFGs to EFGs using Algorithm~II took 9 seconds for all relevant functions. All experiments were carried out on a Windows 8, Intel Core i7 2.40Ghz, 8GB RAM laptop computer.

Table~\ref{table:summary} summarizes for the Linux kernel (v3.12) the number of artifacts:  \emph{LOC} - lines of code, \emph{Srce Files} - source files, \emph{Functions} - functions, \emph{Rlvnt. Func.} - \emph{relevant} functions, \emph{Spin Objs.} - spin lock objects, and \emph{Events} - events of interest to the \emph{spin safe-synchronization} property.

\begin{table}[h]
    \setlength{\tabcolsep}{5pt}
    \renewcommand{\arraystretch}{1.2}
    \scriptsize
    \caption{Program artifacts of the Linux Kernel (v3.12)}
  \centering
    \begin{tabular}{|c|c|c|c|c|c|}
        \hline
        LOC & Srce Files & Functions & Rlvnt. Func. & Spin Objs. & Events \\
        \hline
        11,479,683 &  36,613 & 63,190 & 3,894 & 531 & 8,086\\
        \hline
    \end{tabular}
  \label{table:summary}
\end{table}

\subsection{Experiment I: Reductions from CFG to EFG}
We present the reduction in nodes and edges in going from CFGs to EFGs. Recall that the EFG consists of the event and relevant branch nodes and that the reduction is due to the removal of non-event nodes and irrelevant branch nodes.

Table~\ref{table:dist-cfg-efg} shows the distribution of nodes, edges, and branch nodes for both the CFGs and EFGs for all the \emph{relevant} functions ($F_{\text{Relevant}}$). In assessing graphs with a large number of nodes, we find only 15 EFGs compared to 1,058 CFGs that have ($>30$) nodes, i.e., a reduction of $\sim$99\%. In assessing graphs with a large number of edges, we find only 76 EFGs compared to 1,309 CFGs that have ($>30$) edges, i.e., a reduction of $\sim$94\%. In assessing graphs with a large number of branch nodes, we find only 107 EFGs compared to 597 CFGs that have ($>10$) branch nodes, i.e., a reduction of $\sim$90\%. In assessing straightforward cases for checking feasibility of paths, we find 559 CFGs compared to 1,458 EFGs with no branch nodes, i.e., a 161\% increase.

\begin{table}[ht]
    \setlength{\tabcolsep}{4.5pt}
    \renewcommand{\arraystretch}{1.2}
  \scriptsize
  \caption{CFG and EFG statistics for the 3,894 relevant functions ($F_\text{Relevant}$) in Linux kernel (v3.12)}
  \centering
    \begin{tabular}{c|r|r|r|r|r|r}
    \hline
    Graph & \multicolumn{1}{|c|}{Artifact} & \multicolumn{5}{c}{Distribution} \\
    \hline
    \multirow{6}[0]{*}{\begin{sideways}CFG\end{sideways}} & \multicolumn{1}{c|}{\multirow{2}[0]{*}{Nodes}} & $\leq 5$  & $6 \rightarrow 10$ & $11 \rightarrow 30$ & $31 \rightarrow 50$ & $> 50$ \\
          \cline{3-7}
          & \multicolumn{1}{c|}{} & 185  & 759 & 1,892 & 614 & 444 \\
          \cline{2-7}
          & \multicolumn{1}{c|}{\multirow{2}[0]{*}{Edges}} & $\leq 5$  & $6 \rightarrow 10$ & $11 \rightarrow 30$ & $31 \rightarrow 50$ & $> 50$ \\
          \cline{3-7}
          & \multicolumn{1}{c|}{} & 266   & 661 & 1,658  & 691  & 618\\
          \cline{2-7}
          & \multicolumn{1}{c|}{\multirow{2}[0]{*}{Branch Nodes}} & $= 0$     & $1 \rightarrow 5$ & $6 \rightarrow 10$ & $11 \rightarrow 30$ & $> 30$ \\
          \cline{3-7}
          & \multicolumn{1}{c|}{} & 559  & 1,996  & 742  & 493 & 104 \\
    \hline
    \hline
    \multirow{6}[0]{*}{\begin{sideways}EFG\end{sideways}} & \multicolumn{1}{c|}{\multirow{2}[0]{*}{Nodes}} & $\leq 5$  & $6 \rightarrow 10$ & $11 \rightarrow 30$ & $31 \rightarrow 50$ & $> 50$ \\
          \cline{3-7}
          & \multicolumn{1}{c|}{} & 2,185  & 1,246 & 448 & 14 & 1\\
          \cline{2-7}
          & \multicolumn{1}{c|}{\multirow{2}[0]{*}{Edges}} & $\leq 5$  & $6 \rightarrow 10$ & $11 \rightarrow 30$ & $31 \rightarrow 50$ & $> 50$ \\
          \cline{3-7}
          & \multicolumn{1}{c|}{} & 2,159 & 910  & 749 & 63  & 13 \\
          \cline{2-7}
          & \multicolumn{1}{c|}{\multirow{2}[0]{*}{Branch Nodes}} & $= 0$     & $1 \rightarrow 5$ & $6 \rightarrow 10$ & $11 \rightarrow 30$ & $> 30$ \\
          \cline{3-7}
          & \multicolumn{1}{c|}{} & 1,458  & 2,062 & 267 & 102 & 5 \\
    \hline
    \end{tabular}
    \label{table:dist-cfg-efg}
\end{table}

The reductions from CFGs to EFGs are particularly important for complex CFGs, and especially for CFGs with a large number of  branch nodes. Table~\ref{table:EFGReduction} lists the reductions for the ten functions in $F_{\text{Relevant}}$ with the largest number branch nodes\footnote{A complete comparison of all relevant functions for the spin safe-synchronization property in the Linux kernel (v3.12) is available at~\cite{RESULTS}}. The $P$(\%) columns denote the reduction percentages for nodes, edges, and branch nodes. For example, for function \code{xs\_udp\_data\_ready} the reductions from CFG to EFG are: from 1,101 to 8 nodes, from 1,513 to 11 edges, and from 317 to 4 branch nodes.

\begin{table}[ht]
    \setlength{\tabcolsep}{2pt}
    \renewcommand{\arraystretch}{1.2}
  \centering
  \caption{A comparison of CFG vs. EFG}
  \scriptsize
    \begin{tabular}{l|c|c|c||c|c|c||c|c|c}
    \hline
    \multicolumn{1}{c|}{\multirow{2}[0]{*}{Function Name}} & \multicolumn{3}{c||}{Nodes} & \multicolumn{3}{c||}{Edges} & \multicolumn{3}{c}{Branch Nodes}\\
    \cline{2-10}
    \multicolumn{1}{c|}{} &  \multicolumn{1}{c|}{CFG} & \multicolumn{1}{c|}{EFG} & \multicolumn{1}{c||}{P(\%)} & \multicolumn{1}{c|}{CFG} & \multicolumn{1}{c|}{EFG} & \multicolumn{1}{c||}{P(\%)} & \multicolumn{1}{c|}{CFG} & \multicolumn{1}{c|}{EFG} & \multicolumn{1}{c}{P(\%)}\\
    \hline
    $\text{xs\_udp\_data\_ready}^1$   &  1,101 & 8  & 99.3  & 1,513 & 11 & 99.3  & 317 & 4  & 98.7\\
    $\text{tcp\_v4\_err}^2$           &  1,024 & 7  & 99.3  & 1,400 & 9  & 99.4  & 287 & 3  & 99.0\\
    $\text{udpv6\_queue\_rcv\_skb}^2$ &  838   & 17 & 98.0  & 1,153 & 28 & 97.6  & 244 & 12 & 95.1\\
    $\text{udp\_queue\_rcv\_skb}^2$   &  838   & 16 & 98.1  & 1,152 & 26 & 97.7  & 243 & 11 & 95.5\\
    $\text{tcp\_v6\_rcv}^2$           &  732   & 24 & 96.7  & 999   & 41 & 95.9  & 205 & 18 & 91.2\\
    $\text{tcp\_v4\_rcv}^2$           &  731   & 24 & 96.7  & 998   & 41 & 95.9  & 205 & 18 & 91.2\\
    $\text{tcp\_v6\_err}^2$           &  720   & 6  & 99.2  & 984   & 7  & 99.3  & 203 & 2  & 99.0\\
    $\text{tcp\_recvmsg}^2$           &  583   & 41 & 93.0  & 790   & 75 & 90.5  & 173 & 35 & 79.8\\
    $\text{tcp\_v4\_conn\_request}^2$ &  605   & 18 & 97.0  & 822   & 32 & 96.1  & 170 & 16 & 90.6\\
    $\text{tcp\_close}^2$             &  601   & 9  & 98.5  & 815   & 10 & 98.8  & 167 & 2  & 98.8\\
    \hline
    \hline
    \multicolumn{1}{r}{Spin objects:} & \multicolumn{9}{l}{ 1: rpc\_xprt.transport\_lock,  2: sock.sk\_lock.slock}\\
    \hline
    \hline
    \end{tabular}
  \label{table:EFGReduction}
  \vspace{-6pt}
\end{table}

\subsection{Experiment II: Manual Verification}
\label{manual}
For this experiment, we randomly selected 400 locking events from the Linux kernel (v3.12) and asked two analysts\footnote{Two PhD students with $\sim$ 9 years of experience in programming.} to manually verify the spin safe-synchronization property for those events. The analysts were asked to report: 1) the analysis time for each event, 2) the conditions (branch nodes) that were used to decide on a path feasibility in case of a path that has a locking event $e_1(p)$ is not followed by an unlocking event $e_2(p)$ (i.e., violating path), and 3) ``buggy'' or ``safe'' verdict for each locking event where buggy means that the violating path is feasible.

We conducted the experiment in two rounds, where in each round we asked the analysts to verify 200 events. In the first round $R1$: the first analyst $A1$ to use only CFGs where the other analyst $A2$ to use EFGs. In the second round $R2$: $A1$ to use EFGs and $A2$ to use CFGs. The analysts were given the following information:
\begin{itemize}
    \item The function $f_i$ and the line number for the locking event $e_j(p)$ that needs to be verified.
    \item The set of all relevant functions for the analysis of the locking event $e_j(p)$.
    \item The nodes that correspond to the events in $\mathcal{E}_p$ are highlighted in the corresponding CFGs and EFGs. All the branch nodes are diamond-shaped.
\end{itemize}

In case of a path feasibility check, the analyst calculates the Boolean combination, i.e., AND($\wedge$), OR($\vee$), NOT($\neg$), of conditions which must be true for the path to be executed. Then, he checks the satisfiability of that combination. For the sake of simplicity, the analyst can infer the correlation between conditions \emph{only} from the information within the graphs assigned to him.

Table~\ref{table:controlled-experiment-results} shows the analysis times for each analyst in each round along with the number of safe and buggy events. Also, it shows the number of times path feasibility (\code{Feas.}) is conducted. Both analysts came up with identical verdicts for safe and buggy and reported exactly the same set of conditions/branch nodes used for checking feasibility. $A1$ in $R2$ and $A2$ in $R1$ reported that EFGs were enough to perform the verification and the relevant branch nodes were \emph{sufficient} for checking feasibility. The number of feasibility checks varies when using CFGs and EFGs. This occurs if multiple paths in the CFG, that are checked for feasibility, happen to be equivalent, hence they are represented by one path (event trace) in the corresponding EFG.

Moreover, both analysts reported a new bug~\cite{LinuxBug} that has been accepted by the Linux community and it turned out to be spanning multiple kernel versions.

\begin{table}[ht]
    \setlength{\tabcolsep}{2pt}
    \renewcommand{\arraystretch}{1.2}
  \centering
  \caption{Controlled Experiment Results}
  \scriptsize
    \begin{tabular}{c||c|c|c|c||c|c|c|c}
    \hline
    \multicolumn{1}{c||}{\multirow{2}[0]{*}{Analyst}} & \multicolumn{4}{c||}{Round 1 ($R1$)} & \multicolumn{4}{c}{Round 2 ($R2$)}\\
    \cline{2-9}
         & Safe & Buggy & Feas. & Time       & Safe & Buggy & Feas. & Time      \\
    \hline
    $A1$ &  200 &   0   &  49   & 10hr 44min &  199 &    1  &  21   & 2hr 49min \\
    \hline
    $A2$ &  200 &   0   &  18   & 3hr 16min  &  199 &    1  &  58   & 11hr 22min\\
    \hline
    \end{tabular}
  \label{table:controlled-experiment-results}
    \vspace{-6pt}
\end{table}

\section{An EFG Application Study}
\label{case-study}
\begin{figure}[ht]
    \centering
    \includegraphics[width = 0.48\textwidth]{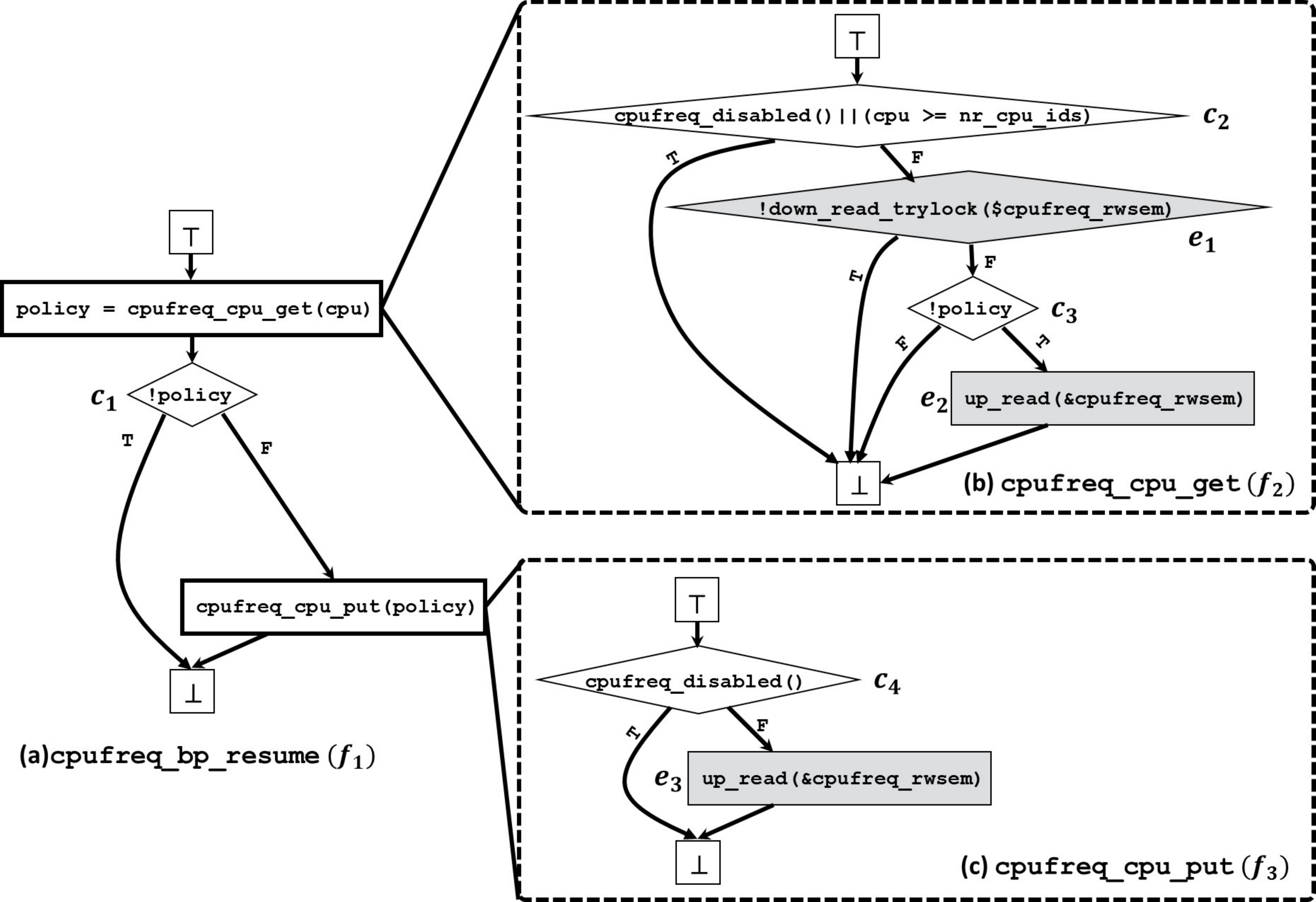}
    \caption{An inter-procedural verification for the locking event in function \code{cpufreq\_cpu\_get}}
    \label{fig:case-study}
    \vspace{-15pt}
\end{figure}

This study presents a real-world example from the Linux kernel (v3.12) to illustrate the applicability and advantages of using EFGs in an inter-procedural path-sensitive verification. It illustrates the verification of the safe-synchronization property for the \emph{read-semaphore} synchronization object (\code{cpufreq\_rwsem}), by verifying that every \emph{locking} event (\code{down\_read, down\_read\_trylock}) is always succeeded by an \emph{unlocking} event (\code{up\_read}) on every feasible execution path. The study involves three functions: \code{cpufreq\_bp\_resume} ($f_1$), \code{cpufreq\_cpu\_get} ($f_2$), and \code{cpufreq\_cpu\_put} ($f_3$). Figure~\ref{fig:case-study} captures the relevant functions for the analysis, and it shows that $f_1$ first calls $f_2$ and then calls $f_3$. The first advantage of using EFGs to perform the analysis is that EFGs are \emph{simpler} and more \emph{compact} than their CFG counterparts. The CFG statistics are given in Table~\ref{table:case-study} and the EFGs are shown in Figure~\ref{fig:case-study}.
\begin{table}[ht]
    \setlength{\tabcolsep}{2.5pt}
    \renewcommand{\arraystretch}{1.2}
  \scriptsize
  \caption{CFG statistics for the application study}
  \centering
    \begin{tabular}{rccc}
        \hline
        Function Name & Nodes & Edges & Branch Nodes\\
        \hline
        cpufreq\_bp\_resume &  16    & 6     & 2    \\
        cpufreq\_cpu\_get   &  17    & 31    & 11   \\
        cpufreq\_cpu\_put   &  6     & 6     & 1    \\
        \hline
    \end{tabular}
  \label{table:case-study}
  \vspace{-10pt}
\end{table}

In $f_2$, there exist four paths corresponding to the four equivalence classes (i.e., unique event traces) of paths in the CFG. Three of those equivalence classes contain $e_1$ that needs to be verified. In the Linux kernel, function (\code{down\_read\_trylock}) returns \code{1} if the lock occurs, otherwise \code{0}. In this example the $e_1$ event evaluates to \code{true} if the lock does not occur; otherwise, it evaluates to \code{false}. That means that, of the three equivalence classes containing $e_1$, the equivalence class that contains the \code{true} branch is \emph{infeasible} when the lock occurs. Hence, the paths that contain this branch do \textbf{not} require verification. Now, for the other two equivalence classes that contain the \code{false} branch of $e_1$, those paths must have an unlocking event (\code{up\_read}). In $f_2$, the paths that contain the \code{true} branch of the branch node $c_3$ contains unlocking event $e_2$. That means that event $e_1$ is verified on those paths. However, on the path that contains the \code{false} branch of $c_3$ (path $A$), there is no unlocking event. Looking at the EFG of $f_2$ alone, one can conclude that this path $A$ contributes to a violation with respect to the tested property. However, the path $A$ locking event $e_1$ is verified with unlocking event $e_3$ in function $f_3$ via function $f_1$. Here is how this is accomplished:

In $f_1$, the path that corresponds to the \code{false} branch of $c_3$ (path $A$) is inflated into two paths: path $B$ that goes via the \code{true} branch of $c_1$, and path $C$ that goes via the \code{false} branch of $c_1$. One can conclude that path $B$ contributes to a bug since no unlocking event occurs and the function returns. However, path $B$ turns out to be infeasible if path $A$ in $f_2$ is taken. This is true because: (1) path $A$ is the path taken via the \code{false} branch of $c_3$; that means $c_3$ evaluates to \code{false}, (2) $c_1 = c_3$ and the variable \code{policy} is returned from $f_2$ to $f_1$. Hence, $c_1$ must evaluate to \code{false} if path $A$ is taken. This means that path $B$ is infeasible when path $A$ is taken. In other words, path $B$ has no verification need. That leaves path $C$ that is feasible and should have an unlocking event.

In $f_1$, path $C$ calls $f_3$. In $f_3$, path $C$ is inflated into paths $D$ and $E$ that traverses the \code{true} and \code{false} branches of $c_4$, respectively. Using reasoning like that for path $B$, one can conclude that path $D$ contributes to a bug. However, path $D$ is infeasible when path $A$ is taken. This is true because path $A$ is the path taken via the \code{false} branch of $c_3$; that means that it also goes via the \code{false} branch of $c_2$. Hence, $c_2$ evaluates to \code{false} in other words, (\code{cpufreq\_disabled()}) must return \code{false}, so $c_4$ evaluates to \code{false} if path $A$ is taken. This means that path $D$ is infeasible when path $A$ is taken. In other words, path $D$ needs no verification. That leaves us with one path, $E$, that is feasible and has an unlocking event. We can finally conclude that there is no violation for the safe-synchronization property and that $e_1$ in function $f_2$ is verified by $e_2$ in $f_2$ (intra-procedural) and by $e_3$ in $f_3$ via $f_1$ (inter-procedural).

This case study shows that: 1) EFGs can be used to perform both intra- and inter-procedural path-sensitive analyses efficiently and accurately. Also, 2) the branch nodes contained in the EFG are the relevant branch nodes for the events and, in determining the feasibility of paths. In this case study, the number of branch nodes are reduced to 5 compared to 14 branch nodes in the corresponding CFGs.
%This case study illustrates the following important points:
%\begin{itemize}
%    \item EFGs can be used to perform both intra- and inter-procedural path-sensitive analyses efficiently and accurately.
%    \item The function $f_2$ by itself appears to be buggy but, given the context of three functions, it becomes clear that $f_2$ is not buggy.
%    \item The branch nodes contained in the EFG are the relevant branch nodes for the events and, in determining the feasibility of paths. In this case study, we have reduced the number of branch nodes to 5 compared to 14 branch nodes in the corresponding CFGs.
%\end{itemize}

\section{Related Work}

%Our Approach
Event-flow graphs are inspired by the work done by Neginhal \emph{et al.}~\cite{Srinivas2006}. They developed the C-Vision tool that introduced the notion of \emph{event view}. C-Vision reductions are based on user-input to determine irrelevant nodes/edges to be removed. There is no algorithmic notion to compute the compact CFG. However, this paper provides a linear-time algorithm to compute the event-flow graph with regard to the given events of interest to the property being analyzed. EFGs enable efficient path-sensitive analyses, and can complement existing analysis techniques that have been researched for path-sensitive analyses.

%Execution Paths Explosion
Many model-checking techniques~\cite{jhala2009software} were developed to verify whether a program's model meets a given property specified in temporal logic. These techniques emphasize precision and accuracy while sacrificing scalability. Many of these techniques explore all paths and result in state space explosion problem~\cite{clarke2004tool, balakrishnan2008slr, harris2010program, dillig2008sound, dor2004software}.

\remove{The light model-checking path-sensitive technique~\cite{Burch90} was developed to verify relatively small finite-state systems such as hardware protocols, where essentially every execution path in the program is symbolically analyzed. Many other heavyweight model-checking techniques enumerate all paths and result in exponential search space~\cite{clarke2004tool, balakrishnan2008slr, harris2010program, dillig2008sound, dor2004software}. In~\cite{le2008marple, bodik1999path} all paths that have demand-driven events corresponding to possible vulnerabilities are only enumerated, however, the number of enumerated paths can become very large.}

%What we do to handle this?
Our approach deals with the problematic exponential number of paths by forming equivalence classes of CFG paths, and analyzing one path from each equivalence class. Our empirical evaluation on the Linux kernel shows that EFGs achieve significant reductions in terms of nodes, edges, and branch nodes, especially for complex CFGs with large numbers of paths and branch nodes. Thus, performing efficient model-checking analyses using EFGs instead of CFGs can be quite beneficial.

%Infeasible Paths Detection
Another line of research focuses on identifying and eliminating infeasible paths before analysis is performed. \cite{bodik1997refining} claimed that 9-40\% of the paths in many programs can be statically identified as infeasible paths. Goldberg \emph{et al.}~\cite{goldberg1994applications} have applied theorem-proving and Ngo and Tan~\cite{ngo2007detecting} have proposed a heuristic approach to identify infeasible paths. Vojdani \emph{et al.}~\cite{vojdani2009goblint} applies the concept of global invariants to deal with the exponentially-large number of paths eliminating infeasible paths. Holley and Rosen~\cite{holley1981qualified} have introduced qualified data-flow analysis to distinguish infeasible paths from the remaining paths. Other researchers have used symbolic evaluation to detect infeasible paths~\cite{navabi2010path, ball2001bebop, xu2008path, dillig2008sound}.

%What we do to handle this?
%Detecting infeasible paths is expensive as it relies on checking the satisfiability of conditions along a path. In our approach, we simplify the detection of infeasible paths by minimizing the number of branch nodes on a path to be checked. This is done by eliminating all irrelevant branch nodes and keeping only the relevant ones. The correlation between relevant conditions can be computed via constant prorogation~\cite{wegman1991constant} or global value numbering~\cite{click1995global} as in~\cite{cho2013}.

Detecting infeasible paths is expensive as it relies on checking the satisfiability of conditions along a path. EFGs can minimize computation for checking path feasibility. As shown in our empirical evaluation of the Linux kernel (Section~\ref{manual}), the relevant branch nodes for forming equivalence classes were also \emph{sufficient} for checking feasibility.

%Execution Effects Merge
Another challenge posed by precise path-sensitive analysis is the separation of execution effects/impact along different paths. Many heuristics schemes are aimed at achieving partial path-sensitive solutions that selectivity join or separate the effects of using different paths using logical disjunctions~\cite{das2002esp, dhurjati2006path, fischer2005joining, mauborgne2005trace, sankaranarayanan2006static, balakrishnan2008slr, dillig2008sound}. Other approaches~\cite{vojdani2009goblint, wang2009improved, dor2004software, jaffarpath, das2002esp, cui2013verifying} determine the relevancy of a branch node through analyzing each individual execution path branching from a branch node. This process requires computation proportional to the number of execution paths.

%The hot-graph path (HPG)~\cite{ammons1998improving} has been proposed to construct a new CFG in which each HPG path is duplicated to eliminate control-flow merging along hot paths. Thakur \emph{et al.}~\cite{thakur2008comprehensive} eliminates merge points by restructuring a program's CFG into a new CFG using a greedy heuristic approach. This incurs an 80\% average increase in the original CFG size.

%BDT to BDD
The Binary Decision Diagram (BDD)~\cite{Akers78} has been used in different contexts of program analysis as a way to reduce the explosion of state space~\cite{ball2001bebop, manevich02, Whaley2004}. The Binary Decision Tree (BDT) to BDD reduction has been also used for path-sensitive analysis~\cite{Xie07, zhang2004}. EFGs can be used in place of BDT to BDD reduction. While EFGs constitute a general technique for simplifying boolean formulas, EFGs have some advantages over BDDs for path-sensitive analysis. The BDT to BDD transformation is similar to those presented in the discussion of Algorithm~I (Section~\ref{algo1}). Unlike BDT to BDD reduction, the EFG transformation achieves further reduction and does not require the input CFG to be acyclic. The EFG transformation deals with cyclic graphs by incorporating a \emph{linear-time} algorithm by Tarjan~\cite{tarjan1972depth} to compute strongly-connected components of a directed graph.

CFG pruning techniques have been proposed in~\cite{murali2007, cho2013} to overcome the computational complexity of exploring all paths. EFGs can complement their techniques as the EFG transformation achieves further reduction in the graph size. Other pruning techniques have been introduced by Choi \emph{et al.}~\cite{choi1991automatic} and Ramalingam~\cite{ramalingam1997sparse} to optimize data-flow graphs. While there is some commonality, those techniques are not well-suited for path-sensitive analyses; the equivalence relation -defined by~\cite{choi1991automatic, ramalingam1997sparse}- is defined with regard to data-flow analysis problems. This equivalence relation is different from the one defined by EFG. Path-sensitive analysis requires preserving the unique event traces and that will not be achieved by the cited techniques.

%We greatly acknowledge the sparse evaluation graph (SEG) by Choi et al. and the compact evaluation graph (CEG) by Ramalingam as efficient techniques to optimize data-flow graphs. While there is some commonality, those techniques are not well-suited for path-sensitive analyses; the equivalence relation -defined by SEG and CEG- is defined w.r.t data-flow analysis problems. This equivalence relation is different from the one defined by EFG. Path-sensitive analysis requires preserving the unique event traces and that will not be achieved by the cited techniques. 
\section{Conclusion}
Efficient and accurate path-sensitive analyses pose challenges of: (a) analyzing the exponentially-increasing number of paths in a CFG, and (b) checking feasibility of paths in a CFG. This paper presents a technique that uses \emph{equivalence classes} of CFG paths to address these challenges. We introduce the notion of \emph{event-flow graph} (EFG) and present a \emph{linear-time} algorithm to compute equivalence classes by compacting a CFG into an EFG. Each path in the EFG represents an equivalence class of paths in the CFG. We show that it is enough to perform path-sensitive analyses only on the equivalence classes produced by an EFG rather than on all the individual paths in the CFG.

Our empirical evaluation on the Linux kernel (v3.12) shows that using EFGs can significantly improve efficiency of path-sensitive analyses.
%Our evaluation was done using C-Atlas, a computational modeling platform from EnSoft. After an initial setup time of about 34 minutes, the conversion from CFGs to EFGs for the 3,894 functions relevant to the events of interest for the spin safe-synchronization property used as a test case took 9 seconds.
Moreover, our controlled experiment results show that EFGs are human comprehensible and compact compared to their corresponding CFGs as they impose fewer paths to verify and fewer branch nodes for feasibility checking. We illustrated an application of EFGs to perform intra- and inter-procedural path-sensitive analyses.

For future work, we are currently developing a verification framework for verifying the safe-synchronization property and analyzing memory leaks in the Linux kernel. The framework leverages the EFG-based path-sensitive analyses to enable developing a sound verification framework that can scale well to large systems such as the Linux kernel.

%Our future work is to develop a verification framework using EFG-based path-sensitive analysis for a variety of verification properties especially the safe-synchronization and memory leak properties.
%The feasibility check can be automated as in~\cite{cho2013} via GVN~\cite{click1995global} or constant propagation~\cite{wegman1991constant}.

% \fi
\IEEEpeerreviewmaketitle

\appendix[Theorems' and Corollaries' Proofs]
\label{appendix}

%\begin{theorem}
%\label{theorem-1}
%Given a CFG $G$, a set $\mathcal{E}$ of events, and the equivalence relation $\mathcal{R}_\mathcal{E}$, there is a one-to-one and onto mapping between the equivalence classes of $\mathcal{R}_\mathcal{E}$ and the paths of the EFG $G_{\text{EFG}}$ where each EFG path produces the event trace corresponding to an equivalence class.
%\end{theorem}

\noindent\textbf{Theorem~\ref{theorem-1} Proof.} The equivalence relation $\mathcal{R}_\mathcal{E}$ partitions the CFG paths into equivalence classes such that all paths in an equivalence class have the same event trace, and the CFG paths that are in different equivalence classes have different event traces.

Since $G_{\text{EFG}}$ is the node-induced subgraph of the given CFG $G$ consisting of the event and the relevant branch nodes, it follows that given an EFG path $S$, it will produce an unique event trace $T$ and conversely given an event trace $T$ there will be a unique EFG path $S$ for which the event trace is $T$. Thus, there is a one-to-one and onto mapping between the equivalence classes of $\mathcal{R}_\mathcal{E}$ and the paths of the EFG $G_{\text{EFG}}$.

%\begin{theorem}
%\label{theorem-2}
%Given a \emph{2-event property} $P$, its verification on all CFG paths can be done using the event traces.
%\end{theorem}

\noindent\textbf{Theorem~\ref{theorem-2} Proof.} If property $P$ holds for an object $p$ on all CFG paths then it clearly holds for all corresponding event traces. Therefore, the case we must argue is the one in which property $P$ is violated for object $p$ on a CFG path $S$. Let $T$ be the event trace for path $S$. Path $S$ may pass through many branch nodes. We will argue that only the relevant branch nodes on that path are important in determining the existence of a feasible path with trace $T$. We will argue that there exists a feasible CFG path with trace $T$ if and only if there exists a path $S^\prime$ with trace $T$ that is feasible with respect to the relevant branch nodes on $S$.

If every path with trace $T$ is infeasible with respect to the relevant branch nodes, then all paths equivalent to $S$ are also infeasible, because the addition of irrelevant branch nodes cannot turn an infeasible path into a feasible one. On the other hand, suppose there exists a path $S^\prime$ with trace $T$ that is feasible with respect to the relevant branch nodes. By the definition of irrelevant branch nodes, an equivalence class has paths going through all possible branches at an irrelevant branch node, so if the path $S^\prime$ is not feasible due to having some irrelevant branch nodes we can choose feasible branches at those nodes to construct a new CFG path that is feasible and equivalent to $S^\prime$. Thus, if there exists a path $S^\prime$ with trace $T$ that is feasible with respect to the relevant branch nodes on $S$, then there always exists a feasible CFG path with trace $T$.

Thus, if property $P$ is violated on path $S$, then we have the following: (a) if $S$ is feasible with respect to the relevant branch nodes on $S$, then there is a feasible path in the equivalence class of $S$, and the violation of $P$ is a true positive; (b) if $S$ is not feasible with respect to the relevant branch nodes on $S$, then all paths equivalent to $S$ are also not feasible.

\begin{mydef}
The boundary of a subgraph $S$ in a directed graph $G$, denoted by \emph{boundary($S$)}, is the set of nodes $u \in S$ such that \emph{suc($u$)} $\in$ \emph{suc($S$)}.
\end{mydef}

%\begin{theorem}
%\label{theorem-3}
%Let $G$ be a colored \emph{$T$-irreducible} and acyclic graph. Then for any subgraph $S$ containing non-colored nodes of $G$: $|$\emph{suc($S$)}$| \geq 2$.
%\end{theorem}

\noindent\textbf{Theorem~\ref{theorem-3} Proof.} If a non-colored node $u \in G$ has only one successor then it is eliminated by transformation $T_1$. Thus, since $G$ is \emph{$T$-irreducible}, $|$\emph{suc($u$)}$|\geq2$ for all non-colored nodes $u \in G$. Also, by assumption, $G$ is an acyclic graph. Using these two facts, we will show that every subgraph $S$ has a node with at least two successors outside $S$ and thus $|$\emph{suc($S$)}$| \geq 2$.

Let $P_{v_0 \rightarrow v_n} : (v_0, v_1), (v_1, v_2), \cdots, (v_{n-1}, v_n))$ be a maximal path in subgraph $S$. Since $v_n$ is the terminal node of this maximal path $P$, its successor cannot be another node in $S$ \emph{not} on the path $P$. Also, the successor of $v_n$ cannot be another node on the path $P$ because  $G_c$ is an acyclic graph, so $v_n$ must be a boundary node and all its successors must be outside the subgraph $S$. Since $v_n$ is a non-colored node, $|$\emph{suc($v_n$)}$| \geq 2$. Since $v_n$, a node in $S$, has at least two successors outside of $S$, we have $|$\emph{suc($S$)}$| \geq 2 $. This completes the proof.

%\begin{corollary}
%\label{corollary-1}
%\label{cefg}Let $G$ be a CFG and $G_{\text{cEFG}}$ be the condensed EFG. Then, for any subgraph $S$ containing non-colored nodes of $G_{\text{cEFG}}$, $|$\emph{suc($S$)}$|\geq 2$.
%\end{corollary}

\noindent\textbf{Corollary~\ref{corollary-1} Proof.} Note that the condensed EFG $G_{\text{cEFG}}$ is the graph resulting from step (4) of the EFG construction algorithm. By construction, the condensed graph $G_{\text{cEFG}}$ is a colored \emph{$T$-irreducible} graph. Also, by construction $G_{\text{cEFG}}$ is an acyclic graph. By applying the above theorem to $G_{\text{cEFG}}$ we get the proof of the corollary.

%\begin{corollary}
%\label{corollary-2}
%The graph produced by Algorithm II does not contain any irrelevant branch nodes.
%\end{corollary}

\noindent\textbf{Corollary~\ref{corollary-2} Proof.} By construction, the graph $G_{\text{T-irr}}$ resulting after step (1) of Algorithm II, consists of only event nodes, relevant branch nodes, and the irrelevant branch nodes retained by Algorithm I. We will now argue that all the irrelevant branch nodes will be eliminated when $G_{\text{cEFG}}$ is constructed in step (4) of Algorithm II. According to the definition of irrelevant branch nodes (Definition~\ref{irrelevant-branch-nodes}), a node $c$ is irrelevant if there is a subgraph $S$ that contains $c$, all its branch edges, $S$ has no event nodes, and $|$\emph{suc($S$)}$|= 1$. It follows from this definition and from the corollary~\ref{cefg} that $G_{\text{cEFG}}$ does not contain any irrelevant branch nodes. Thus, the final graph produced by Algorithm II also does not contain any irrelevant branch nodes, because it consists of the nodes in $G_{\text{cEFG}}$ and all the event nodes.

\section*{Acknowledgment}
This research was supported by DARPA under agreement number FA8750-12-2-0126.



%\bibliographystyle{IEEEtran}
\bibliographystyle{IEEEtran}
\bibliography{IEEEabrv,references}

\end{document}